\documentclass[10pt,aps,amsmath,amssymb,prl,twocolumn,superscriptaddress]{revtex4-1}
\usepackage[colorlinks=true,linkcolor=blue,citecolor=blue]{hyperref}%
\usepackage{amsfonts}
\usepackage{dcolumn}
\usepackage{bm}
\usepackage{tikz}
\usepackage{amsmath,amsfonts,amssymb,times,natbib}
\usepackage{multirow, makecell}
\usepackage[standard]{ntheorem}
\usepackage{graphicx}  
\usepackage{graphics}
\usepackage{amssymb}   
\usepackage{siunitx}   
\usepackage{amsmath}
\usepackage{amssymb}
\usepackage{setspace}
\usepackage{float}

\begin{document}
\title{Cloning of Quantum Entanglement}
\author{Li-Chao Peng}
\thanks{L.-C. P. and D. W. contributed equally to this work.}
\affiliation{Hefei National Laboratory for Physical Sciences at Microscale and Department of Modern Physics, \\University of Science and Technology of China,  Hefei,  Anhui 230026,  China}
\affiliation{CAS Center for Excellence and Synergetic Innovation Center in Quantum Information and Quantum Physics, University of Science and Technology of China, Hefei, Anhui 230026, China}
\author{Dian Wu}
\thanks{L.-C. P. and D. W. contributed equally to this work.}
\affiliation{Hefei National Laboratory for Physical Sciences at Microscale and Department of Modern Physics, \\University of Science and Technology of China,  Hefei,  Anhui 230026,  China}
\affiliation{CAS Center for Excellence and Synergetic Innovation Center in Quantum Information and Quantum Physics, University of Science and Technology of China, Hefei, Anhui 230026, China}
\author{Han-Sen Zhong}
\affiliation{Hefei National Laboratory for Physical Sciences at Microscale and Department of Modern Physics, \\University of Science and Technology of China,  Hefei,  Anhui 230026,  China}
\affiliation{CAS Center for Excellence and Synergetic Innovation Center in Quantum Information and Quantum Physics, University of Science and Technology of China, Hefei, Anhui 230026, China}
\author{Yi-Han Luo}
\affiliation{Hefei National Laboratory for Physical Sciences at Microscale and Department of Modern Physics, \\University of Science and Technology of China,  Hefei,  Anhui 230026,  China}
\affiliation{CAS Center for Excellence and Synergetic Innovation Center in Quantum Information and Quantum Physics, University of Science and Technology of China, Hefei, Anhui 230026, China}
\author{Yuan Li}
\affiliation{Hefei National Laboratory for Physical Sciences at Microscale and Department of Modern Physics, \\University of Science and Technology of China,  Hefei,  Anhui 230026,  China}
\affiliation{CAS Center for Excellence and Synergetic Innovation Center in Quantum Information and Quantum Physics, University of Science and Technology of China, Hefei, Anhui 230026, China}
\author{Yi Hu}
\affiliation{Hefei National Laboratory for Physical Sciences at Microscale and Department of Modern Physics, \\University of Science and Technology of China,  Hefei,  Anhui 230026,  China}
\affiliation{CAS Center for Excellence and Synergetic Innovation Center in Quantum Information and Quantum Physics, University of Science and Technology of China, Hefei, Anhui 230026, China}
\author{Xiao Jiang}
\affiliation{Hefei National Laboratory for Physical Sciences at Microscale and Department of Modern Physics, \\University of Science and Technology of China,  Hefei,  Anhui 230026,  China}
\affiliation{CAS Center for Excellence and Synergetic Innovation Center in Quantum Information and Quantum Physics, University of Science and Technology of China, Hefei, Anhui 230026, China}
\author{Ming-Cheng Chen}
\affiliation{Hefei National Laboratory for Physical Sciences at Microscale and Department of Modern Physics, \\University of Science and Technology of China,  Hefei,  Anhui 230026,  China}
\affiliation{CAS Center for Excellence and Synergetic Innovation Center in Quantum Information and Quantum Physics, University of Science and Technology of China, Hefei, Anhui 230026, China}
\author{Li Li}
\affiliation{Hefei National Laboratory for Physical Sciences at Microscale and Department of Modern Physics, \\University of Science and Technology of China,  Hefei,  Anhui 230026,  China}
\affiliation{CAS Center for Excellence and Synergetic Innovation Center in Quantum Information and Quantum Physics, University of Science and Technology of China, Hefei, Anhui 230026, China}
\author{Nai-Le Liu}
\affiliation{Hefei National Laboratory for Physical Sciences at Microscale and Department of Modern Physics, \\University of Science and Technology of China,  Hefei,  Anhui 230026,  China}
\affiliation{CAS Center for Excellence and Synergetic Innovation Center in Quantum Information and Quantum Physics, University of Science and Technology of China, Hefei, Anhui 230026, China}
\author{Kae Nemoto}
\affiliation{National Institute of Informatics, 2-1-2 Hitotsubashi, Chiyoda-ku, Tokyo 101-8430, Japan}
\author{William J. Munro}
\affiliation{NTT Basic Research Laboratories \& NTT Research Center for Theoretical Quantum Physics, NTT Corporation, 3-1 Morinosato-Wakamiya, Atsugi, Kanagawa 243-0198, Japan}
\affiliation{National Institute of Informatics, 2-1-2 Hitotsubashi, Chiyoda-ku, Tokyo 101-8430, Japan}
\author{Barry C. Sanders}
\affiliation{Hefei National Laboratory for Physical Sciences at Microscale and Department of Modern Physics, \\University of Science and Technology of China,  Hefei,  Anhui 230026,  China}
\affiliation{CAS Center for Excellence and Synergetic Innovation Center in Quantum Information and Quantum Physics, University of Science and Technology of China, Hefei, Anhui 230026, China}
\affiliation{Institute for Quantum Science and Technology, University of Calgary, Alberta T2N 1N4, Canada}
\affiliation{Program in Quantum Information Science, Canadian Institute for Advanced Research, Toronto, Ontario M5G 1Z8, Canada}
\author{Chao-Yang Lu}
\affiliation{Hefei National Laboratory for Physical Sciences at Microscale and Department of Modern Physics, \\University of Science and Technology of China,  Hefei,  Anhui 230026,  China}
\affiliation{CAS Center for Excellence and Synergetic Innovation Center in Quantum Information and Quantum Physics, University of Science and Technology of China, Hefei, Anhui 230026, China}
\author{Jian-Wei Pan}
\affiliation{Hefei National Laboratory for Physical Sciences at Microscale and Department of Modern Physics, \\University of Science and Technology of China,  Hefei,  Anhui 230026,  China}
\affiliation{CAS Center for Excellence and Synergetic Innovation Center in Quantum Information and Quantum Physics, University of Science and Technology of China, Hefei, Anhui 230026, China}

\begin{abstract}  
Quantum no-cloning, the impossibility of perfectly cloning an arbitrary unknown quantum state, is one of the most fundamental limitations due to the laws of quantum mechanics, which underpin the physical security of quantum key distribution. Quantum physics does allow, however, approximate cloning with either imperfect state fidelity and/or  probabilistic success. Whereas approximate quantum cloning of single-particle states has been tested previously, experimental cloning of quantum entanglement -- a highly non-classical correlation -- remained unexplored. Based on a multiphoton linear optics platform, we demonstrate quantum cloning of two-photon entangled states for the first time. Remarkably our results show that one maximally entangled photon pair can be broadcast into two entangled pairs, both with state fidelities above 50\%. Our results are a key step towards cloning of complex quantum systems, and are likely to provide new insights into quantum entanglement.
\end{abstract}
\maketitle
\begin{figure}[t]
  \centering
  \includegraphics[width=0.4\textwidth]{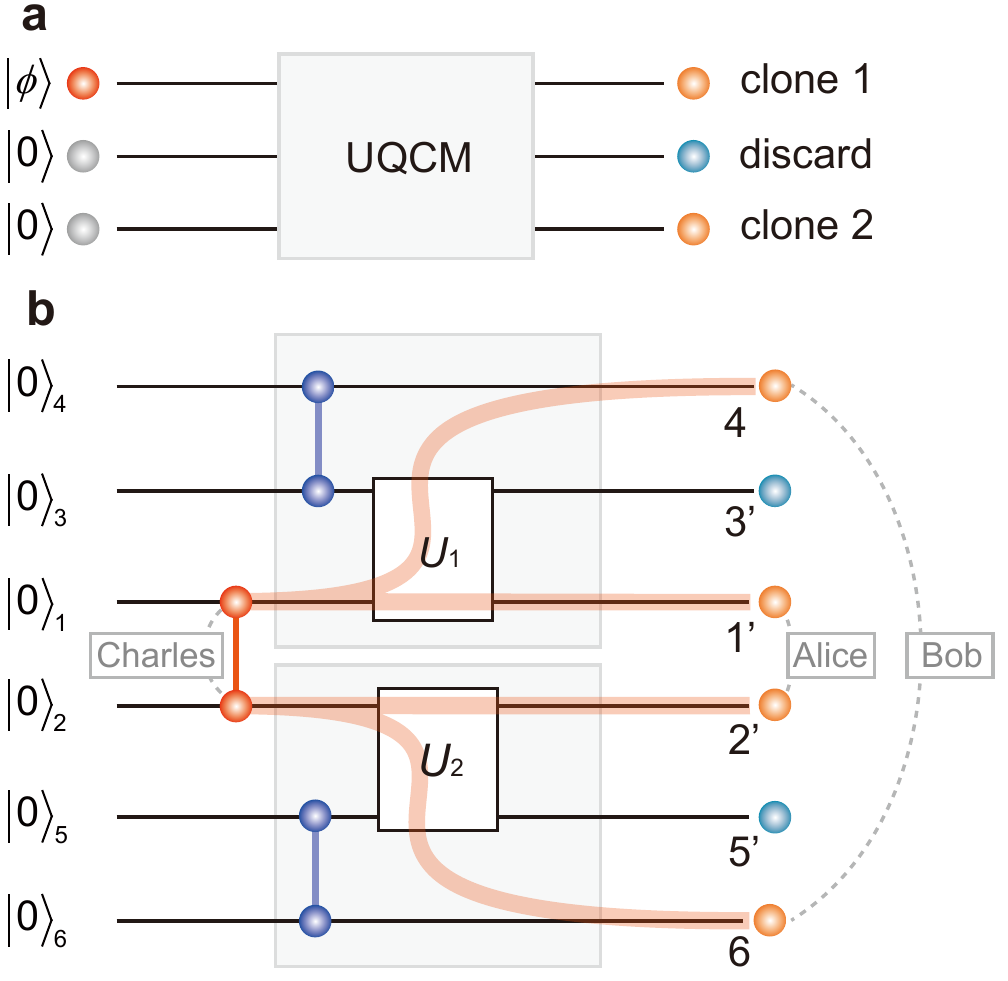}
  \caption{\textbf{Operational principle of our quantum cloning network}.  \textbf{a} depicts a universal single-qubit quantum cloning machine, which generates two approximate clones for an unknown quantum state $|\phi \rangle$ with the support of two ancilla qubits. Next \textbf{b} shows our entanglement cloning network. Charles first prepares his qubits $|0\rangle_1$ and  $|0\rangle_2$ into an entangled state $|\phi \rangle_{12}$ (vertical red line). Next the qubit $1$ ($2$) are fed into a UQCM (grey box) with qubits $|0\rangle_3$ ($|0\rangle_5$) and $|0\rangle_4$ ($|0\rangle_6$). Charles then simultaneously performs quantum cloning on his two-qubit state by enacting unitary operations $\hat{\mathbf{U}}_1$ and $\hat{\mathbf{U}}_2$ on qubits 1, 3 and qubits 2, 5. Consequently, Bob receives the two-qubit pair ($4$, $6$) while Alice gets her two-qubit pair ($1'$, $2'$) respectively.}
  \label{Fig:1}
\end{figure}
``Information is physical" was a profound statement codified by Landauer to mean  that information is not an abstract entity and only exists through a physical representation \cite{Landauer1999}. Classical information can, in principle, be precisely measured, perfectly cloned, broadcast, and deleted. However, quantum information is radically different. Nature prevents us from constructing a quantum machine to produce perfect copies of an unknown quantum state \cite{wootters1982}. 

One fundamental question that naturally arises concerns how much one can extract quantum information of a quantum system from its imperfect copies. To investigate this limitation, various quantum cloning machines (QCMs) that produce approximate copies with non-unity state fidelities  \cite{buzek1996} (defined as the overlap between cloned output states and initial input state) or probabilistic success  \cite{duan1998} have been theoretically investigated and experimentally demonstrated in a variety of systems \cite{scarani2005,fan2014} including single photons \cite{lamas2002,de2004,irvine2004,nagali2009,zhao2005}, nuclear magnetic resonance \cite{cummins2002} and superconducting circuits \cite{yang2019}. In parallel to these fundamental efforts, quantum cloning has also been exploited as a powerful tool for the investigation of the quantum-to-classical transition \cite{de2008}, quantum state estimation \cite{bruss1998}, quantum cryptanalysis \cite{bartkiewicz2013} and complementary \cite{thekkadath2017}.

Quantum correlations are at the heart of this cloning phenomena. An interesting yet experimentally unexplored regime relates to the cloning of quantum entangled states—distant particles in an inseparable state. Theoretically, two pairs of nonlocally entangled pairs can be generated by locally performing quantum cloning on each subsystem comprising one entangled pair -- a technique called entanglement broadcasting \cite{buzek1997,buzek1998,ghiu2005}. Interest in this link between quantum cloning and entanglement lies not only in the extension from cloning of qubits to registers \cite{buzek1998}; profoundly, this link could reveal that entanglement, as a novel quantum resource without any classical counterpart, has the quantum feature of broadcasting, similar to other basic behaviors such as manipulation, control and distribution \cite{horodecki2009}. 

Entanglement cloning is potentially useful for quantum network with multiple functional quantum node that can distribute, control and manipulate entanglement. While quantum teleportation aims to transmit the entangled states faithfully, entanglement broadcasting is a more general concept that seeks to spread the entanglement among multiple parties. Broadcasting entanglement provides an attractive alternative to supply entanglement to quantum networks which currently involve Bell-pair creation between quantum network nodes \cite{wehner2018,van2013} or multipartite entangled states \cite{pirker2019}.

Here we report on the first experimental cloning of quantum entanglement. In Figure\ \ref{Fig:1} we depict the concept of our operational quantum network for entanglement broadcasting with six initialized qubits. It begins by Charles preparing an entangled pure bipartite state $|\phi\rangle_{12}$ on two spatially separated qubits. Charles wants to make a copy of $|\phi\rangle_{12}$ for Bob at a distant location while saving a local transcript for Alice. This is unfortunately prohibited by the no-cloning principle but Charles can instead employ two universal 1-to-2 single-qubit quantum cloning machines (UQCMs) \cite{buzek1996,brub1998}. These UQCMs (depicted  Fig.\ \ref{Fig:1} a) generate two approximate copies of an arbitrary input qubit $|\phi\rangle$ with cloning fidelities independent of the input state. One experimentally viable UQCM protocol employs the Pauli cloning machine \cite{cerf2000a} by performing partial teleportation \cite{radim2004}, instead of the symmetrical QCM, which generates two unnecessary identical clones with the fidelities of two output states under control of a single parameter $R$ (satisfying $0 < R < 1$). In Fig.\ \ref{Fig:1}b, each UQCM is internally equipped with a quantum channel that prepares maximally entangled qubit pairs ($3$, $4$) and ($5$, $6$) of the form $| \Psi^{-} \rangle_{34}=\frac{1}{\sqrt{2}}( | 01\rangle_{34}-| 10 \rangle_{34} )$ and $| \Psi^{-} \rangle_{56}=\frac{1}{\sqrt{2}}( | 01\rangle_{56}-| 10 \rangle_{56} )$ respectively. These are referred to as EPR1 and EPR2. The joint quantum state of the initial system, consisting of two UQCMs and the input bipartite state to be cloned is $| \xi \rangle=| \phi \rangle_{12} | \Psi^{-} \rangle_{34} | \Psi^{-} \rangle_{56}$. 
\begin{figure*}
  \centering
  \includegraphics[width=0.75\textwidth]{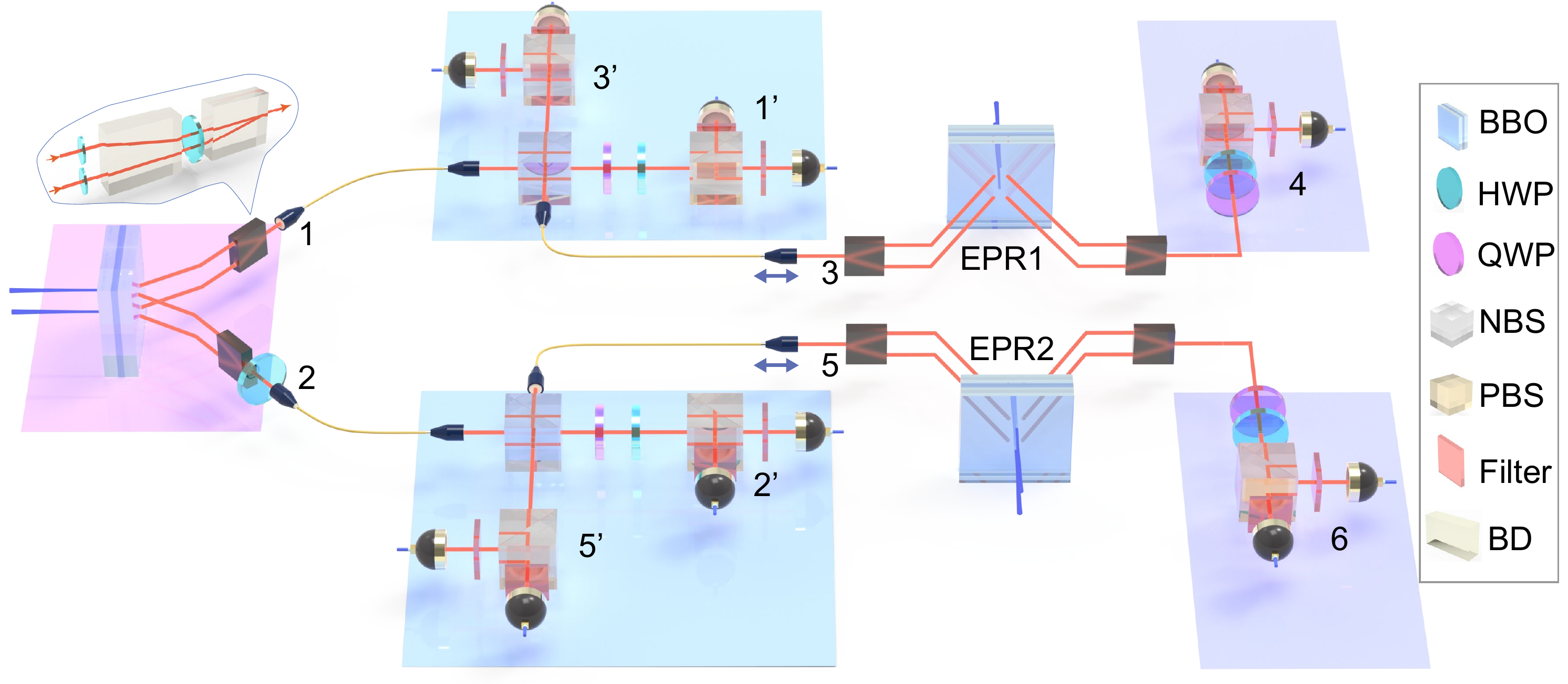}
  \caption{\textbf{Experimental set-up for quantum entanglement cloning.} Here two pulsed Ti:sapphire laser beams (with 775 nm central wavelength and 80 MHz repetition rate) are focused on to a 6.3 mm-thick $\beta$-barium-borate crystal (BBO) to produce photon pairs via SPDC. Optical 4f systems of lenses are inserted before each SPDC source for an optimal beam-waist match of 800 $\mu$m. The photons in spatial modes $1$ and $2$ are combined into one path with BDs and HWPs (inset) and then coupled into single-mode fibers. The HWP in mode $2$ is used to prepare a desired initial state. The photon pairs of EPR1 and EPR2 are prepared in singlet states with similar optical arrangements. Polarization states of photons in modes $1'$, $2'$, $4$ and $6$ are analyzed by using QWPs, HWPs and PBSs. All photons are spectrally filtered by 30-nm band-pass filters. Six-fold coincidences are recorded with a multi-channel coincidence unit. QWP: quarter-wave plate, PBS: polarizing beam splitter, NBS: non-polarizing beam splitter.}
  \label{Fig:2}
\end{figure*}
With our joint initial state $| \xi \rangle$ generated, the next step is to use the two UQCMs. These are locally operated to clone for qubits $1$ and $2$ by performing partial Bell state projections (PBSPs) on the pairs ($1$, $3$) and ($2$, $5$), individually. The transformations labelled $\hat{\mathbf{U}}_1$ and $\hat{\mathbf{U}}_2$  correspond to single-parameter operations of the form
$$
\hat{\mathbf{U}}_1=\hat{\mathbf{U}}_2=\left(\begin{array}{cc}{\sqrt{1-R}} & {i \sqrt{R}} \\ {i \sqrt{R}} & {\sqrt{1-R}}\end{array}\right) \otimes \hat{\mathbf{I}}.
$$
where $\hat{\mathbf{I}}$ is an identity operator for the internal degree of freedom and the $R$ parameterized matrix operates on the spatial modes transforming the whole state of the system as $|\psi\rangle=\hat{\mathbf{U}}_1 \otimes \hat{\mathbf{U}}_2 |\xi \rangle$. Finally modes $3'$ and $5'$ are traced out. At this point, Alice obtains the local two-qubit state $\rho_{1'2'}$ with fidelity $F_{1'2'}$ while Bob receives the distant state $\rho_{46}$ with fidelity $F_{46}$. When $R=1/3$, the two local cloning processes are symmetrical and optimal with the local (distant) state obeying
$$
\rho_{1'2'}=\rho_{46}=\sigma=\operatorname{Tr}_{463'5'}(|\psi\rangle\langle\psi|)=\operatorname{Tr}_{1'2'3'5'}(|\psi\rangle\langle\psi|)
$$ 
where $\sigma$ denotes the reduced density matrices of subsystems after partial traces over the corresponding spatial modes. Interestingly, despite using universal single-state quantum cloning machines here, our approach for bipartite entanglement cloning is state-dependent, which depends on the particular initial bipartite states Charles prepares. One needs to be aware that our scheme enables broadcasting entanglement for a wide range of states including valuable Bell states as well as some non-maximally entangled states with both of the cloning pairs in inseparable states \cite{sup}. 

Turning our attention the experiment itself we show a schematic diagram of our setup in Fig.\ \ref{Fig:2}. We first prepare the initial two qubit input entangled state, which is encoded in the polarization degree of freedom of two individual flying photons. At this preparation stage, two parallel laser beams with the same pump power  are focused on a $\beta$-barium borate (BBO) crystal to generate independent photon pairs via spectral-uncorrelated spontaneous parametric down-conversion \cite{kwait1995,zhong2018}. The frequency-degenerate down-converted photons in identical signal and idler modes are combined into one path by birefringent beam displacers (BDs) and half-wave plates (HWPs). Careful temporal and spatial compensation, by tilting both BDs, enables the preparation of the pair of photons in modes 1 and 2 into our desired initial entangled state $|\phi\rangle_{12}$.  Similarly,  EPR1 and EPR2 are aligned to generate the singlet state $| \Psi^{-} \rangle=\frac{1}{\sqrt{2}}( | H\rangle| V \rangle-| V \rangle | H \rangle )$ for $H$ ($V$) horizontal (vertical) polarization. To suppress higher-order emission noise, our experiments operate at low laser power with an average raw two-fold coincidence count of about $1.5 \times 10^{5}$ per second from each source. Then photons in modes $1$, $2$, $3$ and $5$ are coupled into 2-meter-long single-mode optical fibers (SMFs) and guided for the PBSP operations.  To clone entanglement, photons ($1$, $3$) and ($2$, $5$) must transform unitarily according to $\hat{\mathbf{U}}_1$ and $\hat{\mathbf{U}}_2$, which are implemented on two non-polarizing beam splitters (NBSs) in our linear optical experiment. To achieve perfect non-classical interference, we adjust the path length of photon $3$ ($5$) that guarantees the photons in modes $1$ ($2$) and $3$ ($5$) to arrive at the NBS simultaneously. Furthermore, before each light coupler, photons are spectrally filtered to ensure frequency indistinguishability.

The most critical element of the partial Bell-state measurement in our symmetric UQCM is a non-polarizing beam splitter with reflectivity of  $R=1/3$ for both $H$- and $V$-polarization \cite{radim2004}. While traditional real beam splitters involve imperfect calibration parameters for the reflectivity and transmittance as well as exhibiting asymmetry between different polarizations, our well manufactured beam splitters with polarization-independent coating have reflectivity variation for $H$- and $V$-polarization within $\pm$0.01 according to prior calibration. By adjusting our collimators, we can slightly vary the beam-profile matching and incidence angle for optimal spatial overlap and splitting ratios. Furthermore, we employ Hong-Ou-Mandel-type measurements \cite{hong1987} to estimate the quality of non-classical interferences on our asymmetric beam splitters obtaining a visibility of 0.731$\pm$0.007 (with the maximum ideal visibility being 0.8). This small observed degradation may be from the group-delay dispersion of pump laser in the BBO crystals. Finally, photon polarization in output modes $1'$, $2'$, $4$ and $6$ is analyzed with combinations of half- and quarter-wave plates in conjunction with a polarizing beam splitter. All photons are detected by SMF guided superconducting nanowire single-photon threshold detectors with average detection efficiency of 75\%. 

It is now important to determine the nature of our entanglement cloned states noting that our initial state is prepared in a polarization-entangled state $|\Phi^{+}\rangle_{12}=\frac{1}{\sqrt{2}}(|H\rangle_{1}|H\rangle_{2}+|V\rangle_{1}|V\rangle_{2})$. This choice of initial state is appropriate as the maximally entangled state has the worst cloning fidelity in our state-dependent cloning protocol. Therefore, cloning a Bell state is the most challenging task. We have of course a number of tools available to us to characterize our local and distant states that have been generated. In this situation we primarily want to know if our states are entangled or not and so an entanglement witness seems appropriate. We can define it as $\hat{\mathbf{W}}=\frac{1}{2}\hat{\mathbf{I}}-|\Phi^{+}\rangle\langle\Phi^{+}|$ which is a Hermitian operator with negative expectation value $\operatorname{Tr}[\hat{\mathbf{W}}\tilde{\rho}_{\text{exp}}]<0$ for an entangled state $\tilde{\rho}_{\text{exp}}$ \cite{guhne2009}. Then the expectation value of $\hat{\mathbf{W}}$ can be expressed as
\begin{equation}\nonumber
\langle \hat{\mathbf{W}}\rangle=\operatorname{Tr}[\hat{\mathbf{W}}\tilde{\rho}_{\text{exp}}]=\frac{1}{4}(1-\langle\hat{\sigma}_{x} \hat{\sigma}_{x}\rangle+\langle\hat{\sigma}_{y} \hat{\sigma}_{y}\rangle-\langle\hat{\sigma}_{z} \hat{\sigma}_{z}\rangle),
\end{equation}
where $\hat{\sigma}_{x,y,z}$ are the usual Pauli operators. Thus by measuring the expectation values $\left\langle\hat{\sigma}_{x} \hat{\sigma}_{x}\right\rangle$, $\left\langle\hat{\sigma}_{y} \hat{\sigma}_{y}\right\rangle$ and $\left\langle\hat{\sigma}_{z} \hat{\sigma}_{z}\right\rangle$ of the two-photon output states using correlated local measurements we can determine whether the photon pairs ($1'$, $2'$) and ($4$, $6$) are entangled or not.

\begin{figure}[b]
  \centering
  \includegraphics[width=0.48\textwidth]{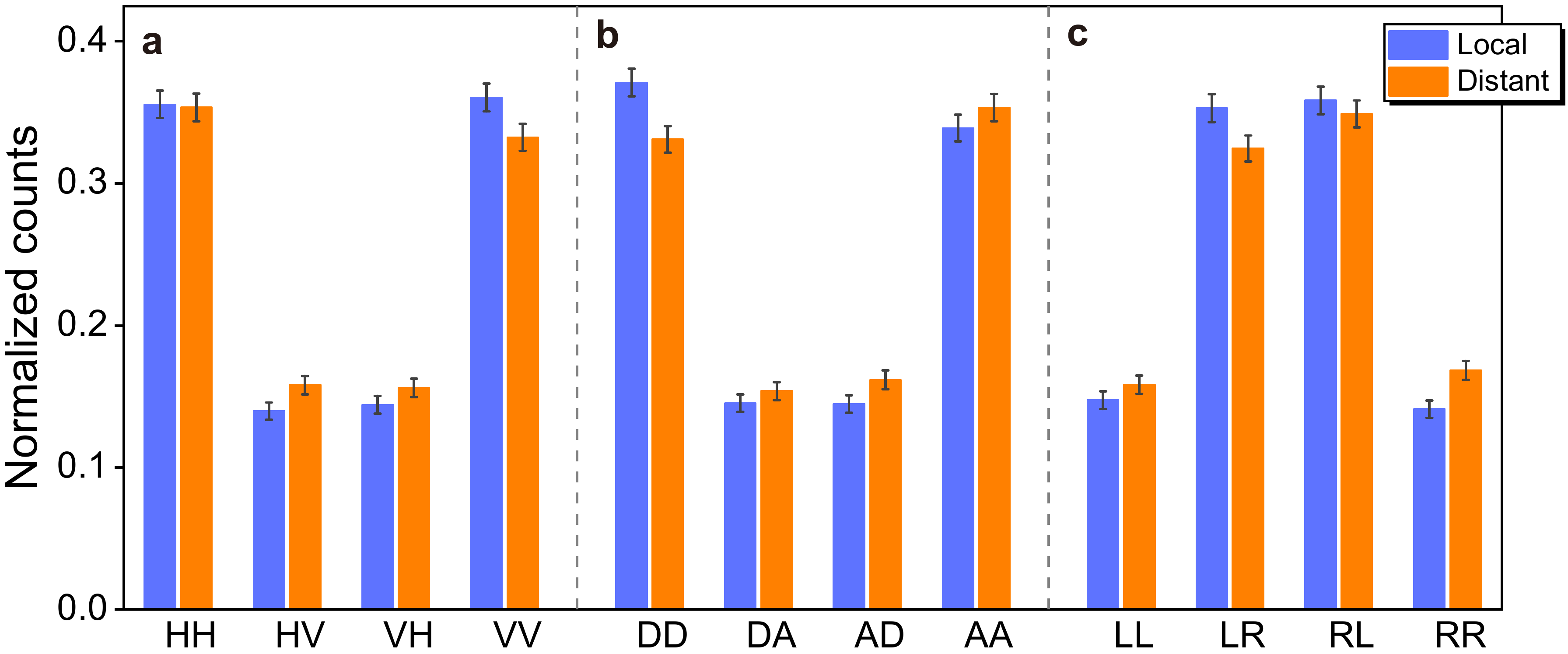}
  \caption{\textbf{Experimental entanglement characterization in three complementary bases}. In \textbf{a} for the linear basis ($|H\rangle/|V\rangle$), the measured expectation value $\langle\hat{\sigma}_{z} \hat{\sigma}_{z}\rangle$ is 0.433$\pm$0.015 (0.372$\pm$0.016) for the local (distant) case. Similarly for \textbf{b,} corresponding to the diagonal basis ($|D\rangle/|A\rangle=|H\rangle\pm|V\rangle$), the measured  $\langle\hat{\sigma}_{x} \hat{\sigma}_{x}\rangle$ is 0.420$\pm$0.015 (0.369$\pm$0.016). Further in \textbf{c,} for the circular basis ($|L\rangle/|R\rangle=|H\rangle\pm i|V\rangle$), the measured $\langle\hat{\sigma}_{y} \hat{\sigma}_{y}\rangle$ is -0.423$\pm$0.015 (-0.347$\pm$0.015). The accumulation time for each setting is 12 hours. The expectation values and error bars are calculated with raw counts.}
  \label{Fig:3}
\end{figure}
In our experiment, we need to measure the entanglement witness of the local (distant) two-photon states  $\rho_{1'2'}$ ($\rho_{46}$)  independently conditioned on the presence of a photon in all the other modes not involved with that state. Given that heralded signal, we then perform polarization measurements on the two photons in our state of interest in  three complementary bases (linear basis ($H$/$V$), diagonal ($D$/$A$) and circular ($L$/$R$). Fig.\ \ref{Fig:3} shows the measured normalized counts of these three measurement settings. We observe that $\langle \hat{\mathbf{W}}\rangle_{1'2'}=-0.069\pm0.007$ while $\langle \hat{\mathbf{W}}\rangle_{46}=-0.022\pm0.007$, which clearly indicates genuine entanglement for both the local and distant two-photon states. Further from these measurements, we can also determine the experimental fidelity $F_{\text{exp}}=\operatorname{Tr}[|\Phi^{+}\rangle\langle\Phi^+|\tilde{\rho}_{\text{exp}}]=\frac{1}{2}-\langle \hat{\mathbf{W}}\rangle$ of our two states of interest. They are 0.569$\pm$0.007 (0.522$\pm$0.007) for the local (distant) states which are close to the theoretical value of 0.583 (0.583) respectively. Thus, our extraction of quantum entanglement for both local and distant states certifies genuine broadcasting of quantum entanglement in our experiment.

\begin{figure}[b]
  \centering
  \includegraphics[width=0.45\textwidth]{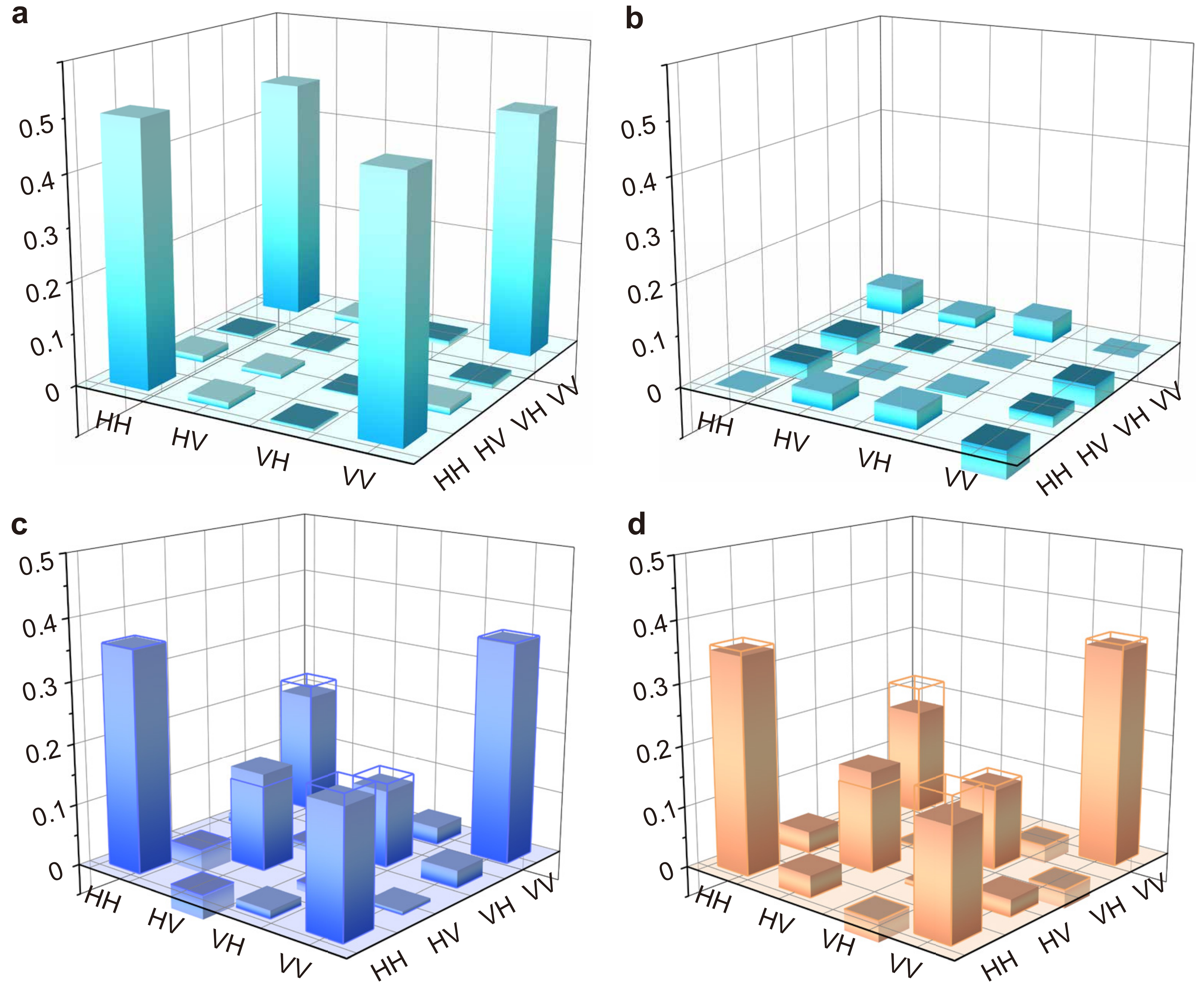}
  \caption{\textbf{Experimental density matrices reconstruction of the initial and cloned states}.\textbf{a} (\textbf{b}), the real (imaginary) part of the density matrices of the initial input bipartite system. In \textbf{c} and \textbf{d} the real parts of the density matrices are shown for the local and distant two-photon cloned state respectively. The elements of imaginary parts are small hence are not shown here. The empty bars illustrate the theoretical values while the solid bars are calculated from experimental raw data.}
  \label{Fig:4}
\end{figure}
Next owing to the intrinsic indistinguishability of cloned states from a symmetric QCM,  our two-photon pairs of ($1'$, $2'$) and ($4$, $6$) should theoretically have identical reduced quantum states $\sigma=\frac{4}{9}|\Phi^{+}\rangle_{12}\langle\Phi^{+}|+\frac{5}{36}\hat{\mathbf{I}}_2 \otimes \hat{\mathbf{I}}_2$. This expression implies that our two states are mixtures of the initial two-qubit maximally entangled state and the maximally mixed state. To quantify experimentally the similarity between these two output states, we can calculate the distance between their two density matrices. To achieve this, we therefore simultaneously perform two-photon correlation measurements on local and distant photon pairs for state tomography with 36 polarization-measurement settings. Composite density matrices are reconstructed using a maximum likelihood algorithm \cite{james2001}. In Figs.\ \ref{Fig:4}a, b, we show the tomographic reconstruction of the initial state with fidelity $F>0.991$ achieved by collecting two-fold coincidences without stray ambient photon subtraction. Figure.\ \ref{Fig:4}c, d show the two cloned output states denoted by $\tilde{\rho}_{1'2'}$ and $\tilde{\rho}_{46}$. The experimental reconstructed state is seen to closely overlap the predicted output state with fidelity of 0.986$\pm$0.006 (0.974$\pm$0.008) for the local (distant) cloning state. Further from the reconstructed density matrices we can also determine the degree of entanglement (concurrence) and mixture (von Neumann entropy) of our local (distant) states. They are $C_{1'2'}=0.146\pm 0.032$ ($C_{46}=0.104 \pm 0.033$) and $S_{1'2'}=1.139 \pm 0.025$ ($S_{46}=1.159 \pm 0.026$) respectively and again clearly showing the presence of entanglement in these states. To quantify the similarity between the local and distant state we can calculate the trace distance $D(\tilde{\rho}_{1'2'}, \tilde{\rho}_{46})=\frac{1}{2} \operatorname{Tr}(|\tilde{\rho}_{1'2'}-\tilde{\rho}_{46}|)=0.197\pm 0.018$ and  Uhlmann state fidelity \cite{jozsa1994} $F(\tilde{\rho}_{1'2'}, \tilde{\rho}_{46})=\operatorname{Tr}(\sqrt{\sqrt{\tilde{\rho}_{1'2'}}\tilde{\rho}_{46}\sqrt{\tilde{\rho}_{1'2'}}})^2=0.948 \pm 0.010$. The error bars are estimated using a Monte Carlo method with noise added according to the Poisson distribution. The small difference between these two measured states could be caused by several experimental issues, for example, residual photon distinguishability between independent SPDC sources and bit-flip errors caused by imperfection of maintaining polarization in glass fiber and in optical elements. Despite such imperfections, remarkably good similarity is seen between the two cloning composite states. We also tested this cloning of quantum entangled states with another
Bell state $|\Psi^+\rangle_{12}=\frac{1}{\sqrt{2}}\left(|H\rangle_{1}|V\rangle_{2}+|V\rangle_{1}|H\rangle_{2}\right)$ and obtained similar result to that of $|\Phi^{+}\rangle_{12}$  (see Fig. \textit{S6}).

\begin{figure}[t]
  \centering
  \includegraphics[width=0.45\textwidth]{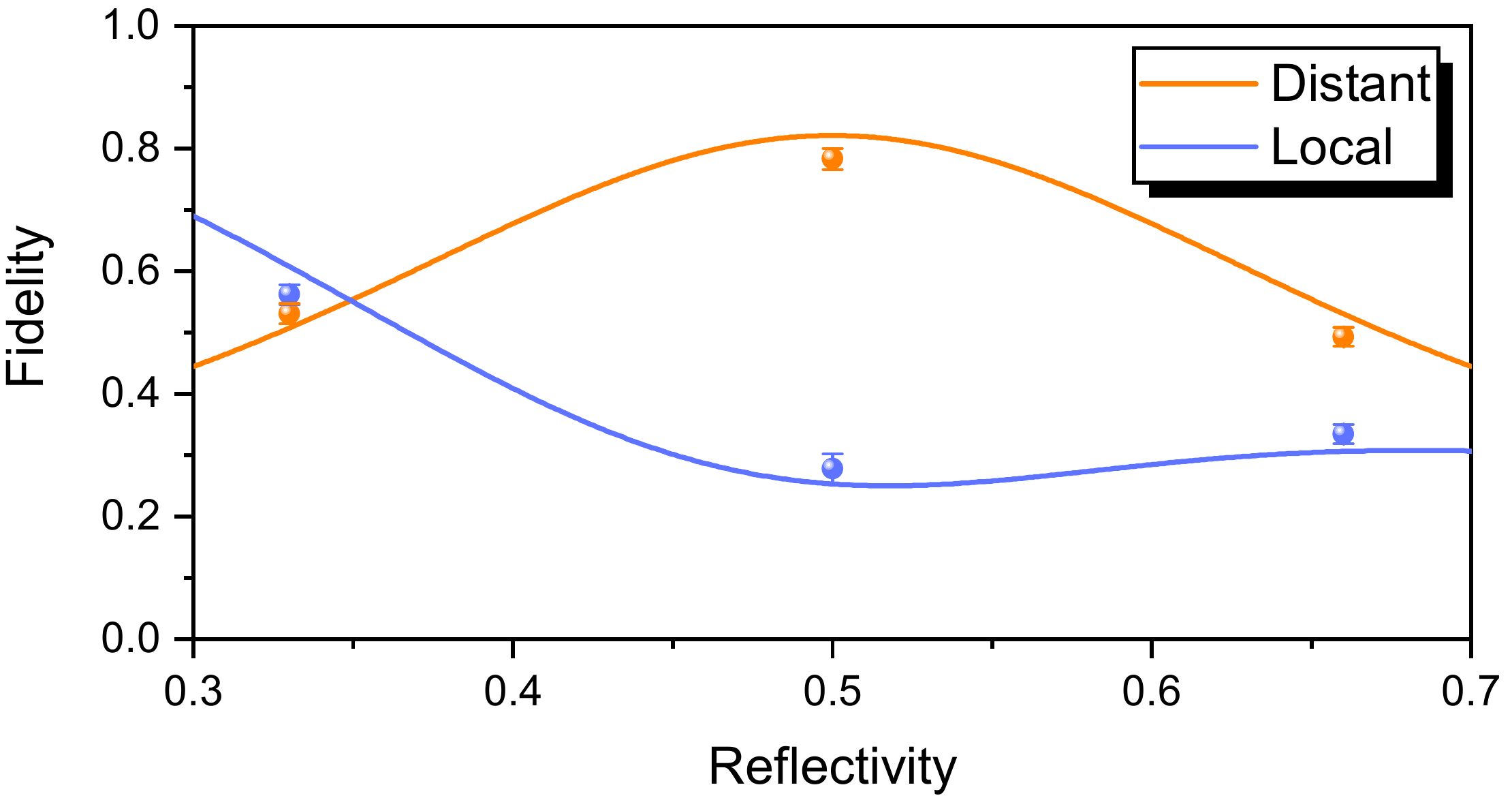}
  \caption{\textbf{Fidelities of the cloned two-photon states for various beam splitter reflectivities.} The dots represent the fidelities determined from the local measurements for $\langle\hat{\sigma}_{k}^{\otimes 4}\rangle(k=x, y, z)$ on photons $1'$, $2'$, $4$ and $6$. Further the solid curves represent the theoretical fidelity of photon pair ($1'$, $2'$) (blue) and ($4$, $6$) (orange) with simulated mode-mismatch noise. The error bars represent one standard deviation.}
  \label{Fig:5}
\end{figure}
Previously we had chosen our PBSPs reflectivity ratio $R=1/3$ such that both the local and distant clones states would have the same fidelities. Of course this need not be the case and so let us exploit asymmetries in PBSPs with various splitting ratios to explore its effect on our local and distant cloned state. In Fig.\ \ref{Fig:5} we show the measured fidelity of the local and distant cloned states (with respect to $| \Phi^{+} \rangle_{12}$) for three different $R$ reflectivities ($1/3$, $1/2$, $2/3$). The measured fidelities are 0.562$\pm$0.017, 0.278$\pm$0.025 and 0.334$\pm$0.016 respectively for photon pair ($1'$, $2'$) and  0.530$\pm$0.017, 0.783$\pm$0.018 and 0.493$\pm$0.015 respectively  for photon pair ($4$, $6$). The solid lines represent our theoretical modeling calculated for experimental parameters involving mode-mismatch noise \cite{sup}. Strong agreement between the experimental measurements and theoretical prediction of our model implies that our setup establishes a convenient and effective tool for quantum entanglement broadcasting. Moreover, competition between the fidelities of the two output states for various beamsplitter reflectivities effectively shows the experimental challenge to simultaneously observe entanglement properties in both local and distant pairs due to the presence of decoherence.

In conclusion, we have for the first time demonstrated cloning of entanglement by producing two entangled pairs each with fidelity exceeding 0.5. Our realization of quantum cloning for an entangled composite system of non-interacting photon pairs not only advances cloning of complex quantum systems but also confirms the broadcasting feature of quantum entanglement. Moreover, it establishes entanglement broadcasting as a primitive along side manipulation, control and distribution for potential applications for entanglement networks, quantum computing and other quantum information protocols.  We expect that our experiment will inspire implementations in other systems (\textit{e.g}., continuous-variable optics \cite{cerf2000b,weedbrook2012}, superconducting circuits \cite{yang2019}) to explore both novel phenomena including quantum-to-classical transitions \cite{de2012} and new applications including quantum-coherent approximate broadcasting \cite{marvian2019,lostaglio2019}.

\begin{acknowledgments}
This work was supported by the National Natural Science Foundation of China, the Chinese Academy of Sciences, the National Fundamental Research Program, and the Anhui Initiative in Quantum Information Technologies.  This Research was also supported in part by the Japanese MEXT Quantum Leap Flagship Program (MEXT Q-LEAP), Grant Number JPMXS0118069605. B. C. S. acknowledges financial support from China’s 1000 Talent Plan.
\end{acknowledgments}

\end{document}